\documentclass[twocolumn,trackchanges,twocolappendix]{aastex631}

\submitjournal{ApJ}
\usepackage{amsmath}
\usepackage{booktabs}
\usepackage{float}
\usepackage{longtable}

\begin{document}
\title{Dark Matter-admixed Rotating White Dwarfs as Peculiar Compact Objects}

\author[0000-0003-2776-082X]{Ho-Sang Chan}
\affiliation{Department of Physics and Institute of Theoretical Physics, The Chinese University of Hong Kong, Shatin, N.T., Hong Kong S.A.R.}
\author[0000-0002-1971-0403]{Ming-chung Chu}
\affiliation{Department of Physics and Institute of Theoretical Physics, The Chinese University of Hong Kong, Shatin, N.T., Hong Kong S.A.R.}
\author[0000-0002-4972-3803]{Shing-Chi Leung}
\affiliation{Department of Mathematics and Physics, SUNY Polytechnic Institute, 100 Seymour Road, Utica, New York 13502, USA}
\affiliation{TAPIR, California Institute of Technology, Mailcode 350-17, Pasadena, CA 91125, USA}

\begin{abstract}
The discoveries of anomalous compact objects challenge our understanding of the standard theory of stellar structures and evolution, so they serve as an excellent laboratory for searching for new physics. Earlier studies on spherically symmetric dark matter-admixed compact stars could explain a handful of anomalies. In this paper, we investigate the observational signatures of dark matter (DM)-admixed rotating white dwarfs and make connections to observed peculiar compact objects. We compute the equilibrium structures of DM-admixed rotating white dwarfs using a self-consistent, two-fluid method, with the DM component being a non-rotating degenerate Fermi gas. We find admixing DM to rotating white dwarfs could: (1) account for some peculiar white dwarfs that do not follow their usual mass-radius relation, (2) allow stable, rapid-rotating white dwarfs that are free from thermonuclear runaway to exist, which could explain some anomalous X-ray pulsars/soft gamma-ray repeaters, and (3) produce universal $I$ (moment of inertia)-Love (tidal Love number)-$Q$ (quadrupole moment) relations that span bands above those without DM admixture, thus providing an indirect way to search for DM in white dwarfs through gravitational-wave detection. To conclude, DM-admixed rotating white dwarfs can account for some peculiar compact objects. Our results suggest a systematic approach to account for unusual compact objects that upcoming surveys could discover.
\end{abstract}


\section{Introduction} \label{sec:intro}
New telescopes such as LISA and LSST are expected to begin operation in the upcoming decade. It is estimated that over 150 million white dwarfs (WDs) could be detected in the final phase of the 10-year LSST survey \citep{Fantin_2020} and that over 25,000 WD binaries could be observed through the LISA telescope \citep{10.1093/mnras/stx1285}. Furthermore, over 10,000 super-luminous Type Ia supernovae per year are expected to be observed by LSST \citep{Villar_2018}, with many more for the total population of Type Ia supernovae. We expect the overwhelmingly large amount of data for WDs and Type Ia supernovae brought by astronomical surveys could contain a considerable number of anomalies, particularly with the help of automatic and robust anomaly detection pipelines \citep{Chan_2022}. These anomalous objects could signify the presence of new physics. One example is the dark matter (DM)-admixed star. DM constitutes more than $95$ \% of the mass in a typical galaxy \citep{2017IJMPD..2630012F} and $26$ \% of the mass-energy budget of the Universe \citep{2012arXiv1201.3942P}. DM could be captured by normal matter (NM) in a region with a high concentration of DM particles \citep{2009ApJ...705..135C, doi:10.1063/1.4868744, 2019Ap&SS.364...24A}. Stars may then contain a DM component. Studies have been made on the effect of admixing DM on the structures and dynamics of compact and exotic objects, such as WD \citep{2013PhRvD..87l3506L}, neutron stars \citep{SANDIN2009278, 2011PhRvD..84j7301L, 2018IJMPD..2750002R}, and quark stars \citep{2016PhRvD..93h3009M}, for which the host object itself is an exotic object to be discovered. \\

These studies show that the effects of DM admixture on compact stars could be significant and observable. These DM-admixed compact stars would have unusual stellar properties, such as masses and radii, and their final fate of evolution would deviate from normal expectations for compact stars without DM admixture. For instance, \citet{2017ApJ...835...33R} discovered that the mass-radius relations of DM-admixed neutron stars agree with some observed anomalous compact objects such as EXO 1745-248, 4U 1608-52, and 4U 1820-30. \citet{2020EPJC...80..544B} showed that the admixture of DM can explain the cooling rate of some pulsars/neutron stars, e.g., PSR B0656+14, PSR B1706-44 and PSR B2334+61, which could not be explained if the popular APR equation of state (EOS) is assumed. \citet{2019ApJ...884....9L} found that the accretion-induced collapse of a DM-admixed WD can explain the formation of some low-mass neutron stars incompatible with the conventional formation path by core-collapse supernovae. \citet{Leung_2015} and \citet{Chan_2021} showed that DM-admixed Type Ia supernovae would produce light curves consistent with some unexplained peculiar supernovae. \citet{PhysRevLett.115.141301} proposed that black-hole forming DM could implode and destroy pulsars, thus explaining the missing pulsar problem near the galactic center. \\

Furthermore, studies on DM-admixed compact stars may lead to constraints on the properties of DM. \citet{2018PhRvD..98k5027G} considered the thermalization of WD cores by DM particles and constrained the DM properties by comparing the measured rate of Type Ia supernovae with those due to the DM heating effect. \citet{PhysRevD.101.115021} constrained the galactic abundance of charged massive relics by studying the old-WD population. \citet{PhysRevLett.107.091301}, \citet{PhysRevD.87.055012}, and \citet{PhysRevD.85.023519} constrained the properties of DM by observation of old neutron stars. \citet{PhysRevLett.113.191301} and \citet{PhysRevD.97.055016} constrained the properties of DM by studying the neutron star population within the galactic center. \cite{2011PhRvD..83h3512K} put constraints on asymmetric DM candidates by investigating DM-admixed WDs and neutron stars. \citet{PhysRevD.105.123010} proposed a method to probe the DM particle mass by measuring the tidal deformability of DM-admixed neutron stars. \citet{2013PhRvD..87e5012B} and \citet{2013PhRvD..87l3507B} constrain the particle mass and annihilation cross section for bosonic DM using DM-admixed neutron stars. These examples show that DM-admixed compact objects could be an important window to uncover new physics through astronomical observations. \\

The non-luminous nature of DM makes it difficult to be detected through conventional telescopes. Its gravitational effects, however, could be indirectly observed through gravitational-wave signatures \citep{PhysRevLett.122.041103}. One important prediction waiting to be confirmed through gravitational-wave detection is the $I-$Love$-Q$ relations. \citet{2013PhRvD..88b3009Y} discovered universal relations that are independent of the EOS among the (all rotationally-induced) moment of inertia $I$, mass quadrupole moment $Q$, and tidal Love number $\lambda_{T}$ of a neutron star. The existence of such universal relations has later been confirmed for WDs \citep{10.1093/mnras/sty1227, roy2020compact, 2020MNRAS.492..978T}. Such universal relations could be due to isodensity self-similarity, an approximate symmetry that emerges for compact stellar objects \citep{2017PhR...681....1Y}. The $I-$Love$-Q$ information of a compact star is imprinted in its gravitational-wave signature \citep{PhysRevD.77.021502, Lau_2010, PhysRevD.81.123016}. Even though the sensitivity of current gravitational-wave detectors does not allow direct independent measurement of the individual $I-$Love$-Q$ numbers, methods for obtaining and testing the universal relations have been proposed \citep{2013PhRvD..88b3009Y, PhysRevD.101.124014}. Similarly, the $I-$Love$-Q$ relations have been applied to constrain extra dimensions \citep{Chakravarti_2020} and gravitational theories beyond relativity \citep{PhysRevD.92.064015, PhysRevLett.126.181101}. Future gravitational-wave measurements of the $I-$Love$-Q$ numbers could provide an excellent laboratory for testing and understanding fundamental physics, including astrophysical DM. \\

The DM admixture adds an extra degree of freedom to the current stellar theory, allowing it to explain both ordinary and anomalous objects. Given that WDs constitute a considerable subset of observed stellar objects and that the DM-admixed model has been successful in accounting for some peculiar compact stars, we believe that DM-admixed WDs deserve more attention. If DM-admixed WDs exist, they would also be promising indirect DM-detecting channels. Furthermore, rotation is essential in studying WD structures and evolution \citep{Yoon_2004, 2005A&A...435..967Y}. Rotating WDs have also been proposed to be progenitors of neutron stars formed by accretion-induced collapses \citep{2005A&A...435..967Y, 2011LRR....14....1F, 2012ApJ...756L...4H} and super-luminous thermonuclear supernovae \citep{2010A&A...509A..75P, 2014MNRAS.445.2340W, 2018A&A...618A.124F}. However, the effect of DM admixture on rotating WDs has never been considered. Moreover, the universal $I-$Love$-Q$ relations are useful for understanding compact-object physics. They have been used to study exotic stars such as dark stars \citep{PhysRevD.96.023005} and neutron stars admixed with quark matter \citep{PhysRevLett.122.061102}. It becomes interesting whether such universal relations continue to hold for DM-admixed rotating WDs. Therefore, we extend previous studies of DM-admixed WDs to rotating WDs to provide predictions to facilitate searches for such objects. The plan of the paper is as follows: Section \ref{sec:method} describes our method of constructing DM-admixed rotating WDs and obtaining the $I-$Love$-Q$ numbers. Section \ref{sec:results} is a summary and discussion of the results. Section \ref{sec:conclusion} concludes our study. The connection of this work with previous work on DM-admixed Type Ia supernovae is discussed in Appendix \ref{sec:barrier}, and the stability of DMRWD is analyzed in Appendix \ref{sec:stab}, Eq. \ref{eqn:randau} for DMRWD is derived in Appendix \ref{sec:2feqs}, and the formation of DMRWD is discussed Appendix \ref{sec:dmawdcreate}.


\section{Methodology} \label{sec:method}
\subsection{Equations of Hydrostatic Equilibrium} \label{sec:hydrostatic}
In this work, we consider a light Fermionic, fully degenerate DM model \citep{PhysRevD.74.063003}. Following the previous studies on DM-admixed WD \citep[DMWD,][]{Chan_2021}, we choose the DM particle mass to be $0.1$ GeV. We assume that the DM component is inherited from a zero-age main-sequence star. The formation of DM-admixed main-sequence stars will be discussed in Appendix \ref{sec:dmawdcreate}. \\

We compute a series of DM-admixed rotating WDs (DMRWDs) by solving the Newtonian hydrostatic equations, including the centripetal force:
\begin{equation}
	\vec{\nabla} P_{i} = - \rho_{i}\vec{\nabla}\Phi + \rho_{i}\omega(s)_{i}^{2}s\delta_{i2}\hat{s}.
\end{equation}
Here, the subscript $i = 1(2)$ denotes the DM (NM) quantities, and $\rho$, $P$, and $\omega$ are the density, pressure, and angular speed of the fluid element, respectively. $s$ is the perpendicular distance from the rotation axis, and $\hat{s}$ is the unit vector orthogonal to and pointing away from that axis. $\omega$ is assumed to be a function of $s$ only. $\delta_{i2}$ is the Kronecker-Delta function, indicating that only the NM is rotating. $\Phi$ is the gravitational potential governed by the two-fluid Poisson equation:
\begin{equation}
	\nabla^{2} \Phi = 4\pi G(\rho_{1} + \rho_{2}).
\end{equation}
The use of the Newtonian framework is justified since the rotation speed and compactness of WDs are small. Following \citet{1985A&A...146..260E}, \citet{1986ApJS...61..479H} and \citet{1994A&A...290..674A}, we can integrate the equation of equilibrium:
\begin{equation}
	\int \frac{dP_{i}}{\rho_{i}} = - \Phi + \delta_{i2}\int \omega(s)_{i}^{2}s ds + C_{i},
\end{equation}
where $C_{i}$ is an integration constant. In particular, following \citet{1986ApJS...61..479H}, we define:
\begin{equation}
	\int \frac{dP_{i}}{\rho_{i}} = H_{i},
\end{equation}
\begin{equation}
	\int \omega(s)_{i}^{2}s ds = -h_{i}^{2}\psi_{i},
\end{equation}
where $H$ is the enthalpy, $\psi$ is the rotational potential, and $h^{2}$ is a constant to be determined. So, the equation of equilibrium can be written in the integral form:
\begin{equation}
\label{eqn:two-fluid}
	H_{i} + \Phi + \delta_{i2}h_{i}^{2}\psi_{i} = C_{i}.
\end{equation}
We assume that the WD is rigidly-rotating, so $\psi_{i} = -s^{2}/2$ and $h_{i} = \omega$. 


\subsection{The Self-Consistent Method} \label{sec:self}
We follow \citet{1986ApJS...61..479H} to adopt an axial symmetric spherical grid. We adopt dimensionless units so that we set the gravitational constant $G$, maximum NM density $\rho_{\text{Max}2}$ and NM equatorial radius $r_{\text{eq2}}$ to be one. The computational domain is described by the radial coordinate $r$ and $\mu = \cos\theta$, where $\theta$ is the polar angle. We divide $\mu$ and $r$ into $N_{\nu}$ and $N_{r}$ equal portions, respectively. Therefore:
\begin{equation}
\begin{aligned}
\begin{cases}
	r_{j} = r_{0}\frac{j - 1}{\text{$N_{r}$} - 1}, & (\text{1 $\leq j \leq$ $N_{r}$}),\\
	\mu_{k} = \frac{k - 1}{\text{$N_{\nu}$} - 1}, & (\text{1 $\leq k \leq$ $N_{\nu}$}).
\end{cases}
\end{aligned}
\end{equation}
Here, $r_{0}$ is the size of the computational domain. In this work, when there is no DM component, we choose $N_{\nu} = 257$, $r_{0} = \frac{16}{15}$, and $N_{r}$ $ = 257$. When a DM admixture occurs, $r_{0}$ and $N_{r}$ are enlarged to accommodate the DM fluid. We compute the gravitational potential in spherical coordinates using the multipole expansion method:
\begin{equation}
\begin{aligned}
	\Phi (\mu, r) = -4\pi G\int_{0}^{\infty}dr'\int_{0}^{1}d\mu'\times\\
	\sum_{n = 0}^{\text{$l_{\text{max}}$}}f_{2n}(r',r)P_{2n}(\mu)P_{2n}(\mu')\rho(\mu', r'),
\end{aligned}
\end{equation}
where $P_{2n}(\mu)$ is the Legendre polynomial, and $l_{\text{max}}$ is the maximum number of moments. We choose $l_{\text{max}} =16$, and:
\begin{equation}
  f_{2n}(r',r) = 
\begin{aligned}
  \begin{cases}
    r'^{2n+2}/r^{2n+1}, & r' < r,\\
    r^{2n}/r'^{2n-1}, & r' > r.
  \end{cases}
\end{aligned}
\end{equation}
To compute the equilibrium structure of a pure NM rotating WD, we need to specify the boundary condition for which $\rho_{2} = 0$. The equatorial boundary is set at $r_{a2} = 1, \theta = \frac{\pi}{2}$, while the axis boundary is set at $0 \leq r_{b2} \leq 1$, $\theta = 0$. The axis-ratio of the NM is defined as $\kappa_{2} = r_{b2}/r_{a2}$. There are 3 degrees of freedom and we need one more parameter to specify the system completely. We choose it to be the radial position of the DM equatorial boundary $r_{a1}$\footnote{In such a case, $r_{b1}$ would be a to-be-determined parameter. One can, of course, choose $r_{b1}$ as the free parameter instead. For rigidly rotating DMRWD, $r_{b1} \approx r_{a1}$, but they could differ if the DMRWD is differentially rotating. However, this scenario will not be discussed in-depth in our current scope.}. We obtain the equilibrium structures by first specifying $\rho_{\rm Max2}, r_{b2}$ and $r_{a1}$ and then solving Eq. (\ref{eqn:two-fluid}) iteratively. We first guess the initial density profiles for the DM and NM and compute $\Phi$. Then, we can get $C_{i}$ from the DM and NM boundaries. After that, we invert Eq. (\ref{eqn:two-fluid}) to obtain the new density profiles from the DM/NM enthalpy. We then update $\Phi$ and $C_{i}$ again. This procedure would continue until the relative changes for $h_{i}$, $C_{i}$ and the maximum of $H_{i}$ for both fluids are less than $10^{-10}$. In general, the DM mass changes during iterations. Therefore, we use the bisection method to vary $r_{b1}$ to obtain the targeted DM mass.


\subsection{Extracting the $I-$Love$-Q$ Information} \label{sec:extract}
We can determine $I$, $\lambda_{T}$ and $Q$ by post-processing the density profiles\footnote{We expect that the relativistic effect of DMRWDs is small, provided that the WD is not rapidly rotating and $\rho_{\text{Max}2}$ is below $\sim 10^{10.5}$ g cm$^{-3}$ We thank the anonymous referee for pointing this out.}. The moment of inertia is given as:
\begin{equation}
\begin{aligned}
	I = I_{1} + I_{2}, \\
	I_{i} = \int \rho_{i} s^{2} d\tau,
\end{aligned}
\end{equation} 
while the mass quadrupole moment is:
\begin{equation}
\begin{aligned}
	Q = Q_{1} + Q_{2}, \\
	Q_{i} = \int \rho_{i} r^{2}P_{2}(\cos\theta) d\tau,
\end{aligned}
\end{equation}
where $d\tau$ is the volume element and $P_{2}(x)$ is the $l = 2$ Legendre polynominal. Following \citet{2017MNRAS.464.4349B}, we scale $I$, and $Q$ by:
\begin{equation}\label{eqn:scaling}
\begin{aligned}
	I \rightarrow \left(\frac{c^{2}}{G}\right)^{2}\frac{I}{M^{3}}, \\
	Q \rightarrow \left(\frac{Mc^{2}Q}{J^{2}}\right). 
\end{aligned}
\end{equation}
Here, $M = M_{1} + M_{2}$ is the total mass, and $J$ is the total angular momentum of the DMRWD. On the other hand, the tidal deformability $\tilde{k}$ is defined as \citep{2017MNRAS.464.4349B}:
\begin{equation} \label{eqn:tidal}
\begin{aligned}
	\tilde{k} = \frac{3 - \tilde{\eta}}{2(2 + \tilde{\eta})}.
\end{aligned}
\end{equation}
$\tilde{\eta}$ is obtained by solving Radau's equation for the unperturbed spherically symmetric configuration \citep{PoissonEric2014GNPR}:
\label{eqn:randau}
\begin{equation}
	\frac{d}{dr}(r\tilde{\eta}) = 6(1 - D(r)(\tilde{\eta} + 1)) - \tilde{\eta}(\tilde{\eta} - 2).
\end{equation}
Here, $D(r) = 4\pi r^{3}\rho(r)/3m(r)$, and $4\pi r^{3}/3m(r)$ is the average density of the enclosed mass at a radial distance $r$. The boundary condition is $\tilde{\eta} (0) = 0$. We show in Appendix \ref{sec:2feqs} the derivation of the two-fluid Radau's equation. To compute Eq. \ref{eqn:tidal}, we would evaluate $\tilde{\eta}$ at the stellar radius $R$ which we take to be the larger of the NM and DM radii. $\lambda_{T}$ is related to $\tilde{k}$ as:
\begin{equation}
    \lambda_{T} = \frac{2R^{5}}{3G}\tilde{k},
\end{equation}
We then scale $\lambda_{T}$ as \citep{2017MNRAS.464.4349B}:
\begin{equation}
	\lambda_{T} \rightarrow \frac{c^{10}\lambda_{T}}{G^{4}M^{5}}.
\end{equation}
Previous studies on the $I-$Love$-Q$ relations assumed slowly rotating WDs, the equilibrium structures of which were computed based on the Hartle formalism. In this work, we reproduce the slowly-rotating limit results by using a $\kappa_{2}$ close to $1$. 


\subsection{Equations of State} \label{sec:eos}
We focus on using the ideal degenerate Fermi gas \citep{1983bhwd.book.....S, 2007coaw.book.....C}, assuming a carbon-oxygen WD with an electron fraction $Y_{e} = 0.5$ for the NM for most of our studies, except that we consider several more EOSs for the NM to explore the $I-$Love$-Q$ relations. These include the Harrison-Wheeler (HW) EOSs \citep{1965gtgc.book.....H} and several parameterized deleptonization formulae to describe the electron fraction at high density, in \citet{1971Ap......7..274A} (ASC), \citet{Liebendorfer_2005} (G15, N13), \citet{Cabez_n_2018} (S15), \citet{2006ApJ...644.1063D} and \citet{2010PhRvD..81d4012A} (VUL).


\section{Results} \label{sec:results}
\subsection{Mass-radius Relations} \label{subsec:peculiar}

\begin{figure} 
	\centering
	\includegraphics[width=1.0\linewidth]{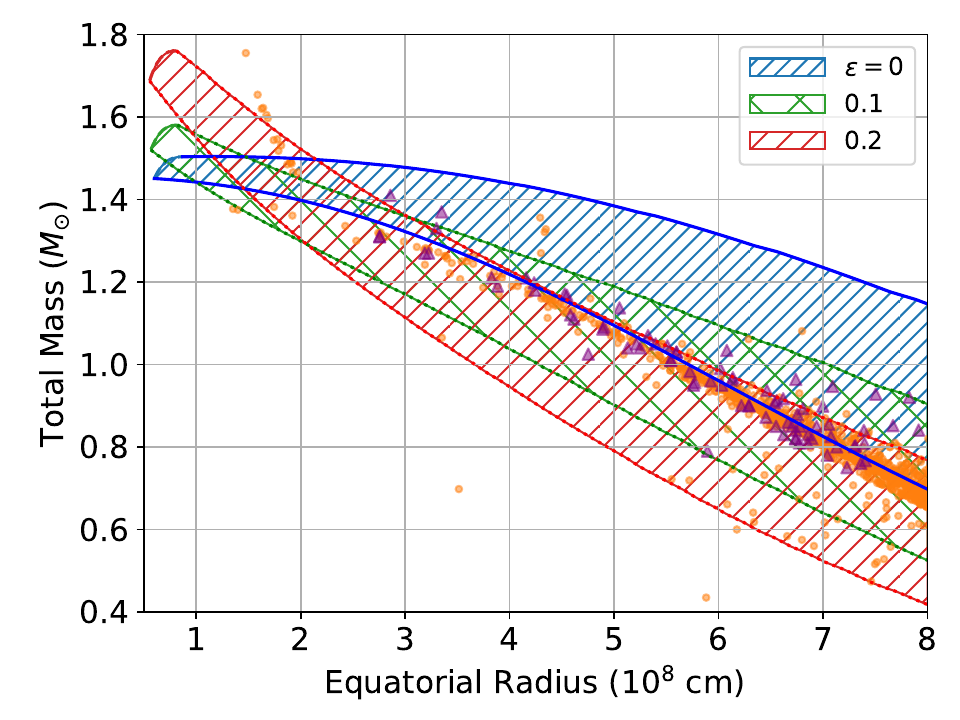}
	\caption{Masses against radii for rigidly-rotating DMRWDs with DM mass fractions $\epsilon = 0, 0.1, 0.2$, indicated by blue, green, and red bands, respectively, spanning possible models from the non-rotating (smaller mass) to the critically rotating (larger mass) limits. We terminate the mass-radius relations at $\rho = \rho_{\beta}$ (c.f. Section \ref{subsec:xray}). We also show data for observed WDs taken from \citet{https://doi.org/10.48550/arxiv.1610.00986} (orange circles) and \citet{2004A&A...420..507N} (purple triangles) as scattered points. We note that the most massive WD confirmed so far has a mass of $1.365$ $M_{\odot}$ \citep{Caiazzo2021}, so that some of the massive WDs from \citet{https://doi.org/10.48550/arxiv.1610.00986} should be interpreted with care. \label{fig:database}}
\end{figure}

The success in using the DM-admixed neutron star model to account for unusual compact objects that do not follow their expected mass-radius relation \citep{2020EPJP..135..637R, 2020EPJP..135..362M, 2021PhRvD.104f3028D, 2021ApJ...922..242L} inspires us to perform a similar analysis for peculiar WDs. We extract observed mass-radius data for WDs from \citet{https://doi.org/10.48550/arxiv.1610.00986} and \citet{2004A&A...420..507N} and show them in Figure \ref{fig:database}. We also append the mass-radius relation in the same Figure for pure NM, rigidly-rotating WDs. Most of the observed WDs agree with the pure NM models\footnote{However, the theoretical line starts to deviate from the majority when the total mass approaches the Chandrasekhar limit, and this could be due to the lack of general relativistic effect and/or the realistic treatment of the equation of state. See, for example, \citet{2013ApJ...762..117B} and \citet{2021ApJ...921..138N}.}. However, there exist some peculiar WDs that are either too light or too massive, lying outside the band of possible pure NM models. \\

\begin{figure} 
	\centering
	\includegraphics[width=1.0\linewidth]{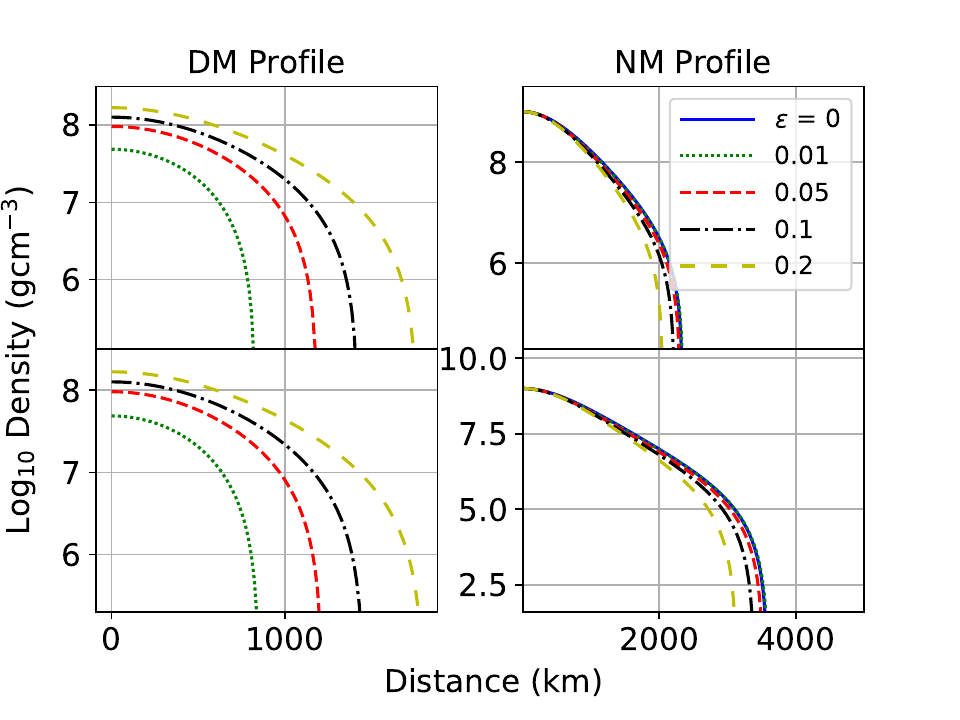}
	\caption{Density profiles of DMRWDs that have different DM mass fractions $\epsilon$. Left panel: DM. Right panel: NM. In each panel, the upper (lower) plot shows the polar (equatorial) density profiles. These DMRWDs all have an NM central density of $10^{9}$ g cm$^{-3}$ and are rigidly rotating at their Keplerian limit. \label{fig:profile}}
\end{figure}

We aim to apply the DMRWD model to explain these anomalies. We compute the mass-radius relations for rigidly-rotating DMRWD with different mass fractions $\epsilon$ and show them in Figure \ref{fig:database}. We find that the admixed DM affects the mass-radius relation in two ways. First, at a given large radius, the total mass is reduced. Second, the Chandrasekhar limit is increased. In particular, the increase in Chandrasekhar limit is not observed in the results by \citet{2013PhRvD..87l3506L}. They assumed heavier DM particle mass in the range of 1-100 GeV. In such a case, the DM component is very compact and affects the global structure of the star significantly. In this work, we assumed the sub-GeV Fermionic DM model. In such a scenario, the DM component is more diffusive and extended (c.f. Figure \ref{fig:profile}), thus producing a less significant effect on the structure of the NM component. We also observe that admixing DM changes the concavity of the mass-radius relation. For instance, at a radius of $10^{8}$ cm, the mass-radius relation changes from concaving upward to downward. The concavity change is present for the mass-radius relation of a pure NM WD when one follows the relation from a larger to a smaller radius. Such a change indicates that the WD is approaching the Chandrasekhar limit. The DMRWD (assuming $0.1$ GeV DM particle mass) has a larger Chandrasekhar limit, and the limit is achieved at a smaller radius (higher density). Thus DMRWDs, say with $\epsilon = 0.2$ and a radius of $10^{8}$ cm, are not yet compact enough to reach the Chandrasekhar limit. That explains why the concavity of the mass-radius relation changes when DM is admixed. \\

\begin{figure} 
	\centering
	\includegraphics[width=1.0\linewidth]{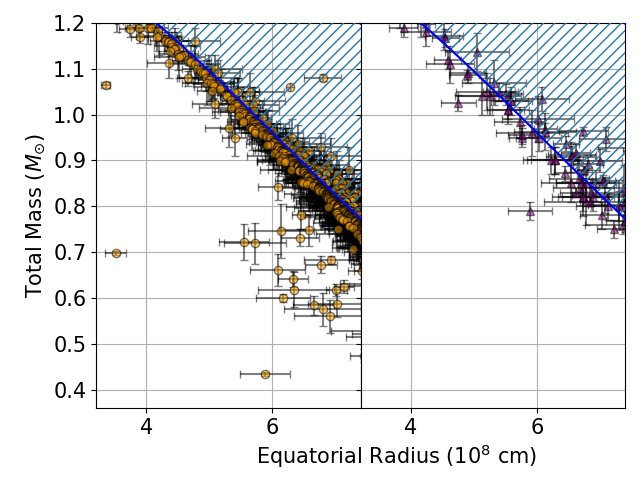}
	\caption{Same as \ref{fig:database}, but for a magnified plot with error bars included. The left (right) panel shows data points extracted from \citet{https://doi.org/10.48550/arxiv.1610.00986} (\citet{2004A&A...420..507N}). Here, we plot only the $\epsilon = 0$ model to facilitate visualization. \label{fig:errorbar}}
\end{figure}

Admixing sub-GeV Fermionic DM to a WD can make it exceptionally light/massive. Figure \ref{fig:database} shows that some of the observed anomalously low-mass/massive WDs lie within the bands of possible models spanned by rigidly-rotating DMRWDs with different $\epsilon$. Note that we sample only data with error bars of less than $10$ \% for both masses and radii and that there are no error bars for very massive WDs, which cast doubts on whether they are actually super-Chandrasekhar. We still include these suspicious data points to show the capability of our model, but readers should interpret these WDs with caution. We magnify Figure \ref{fig:database} to show anomalous low-mass WDs with error bars in Figure \ref{fig:errorbar}. We observe that these anomalous WDs deviate from the pure NM line (blue), which the error bars cannot cover. Such a result shows that alternative explanations are essential to understand these anomalies. To conclude, the DMRWD model could account for WDs with anomalous mass-radius properties.\footnote{When the DMRWD approaches the Chandrasekhar limit, the DM component surrounds the NM component. As such, the measured mass might vary between the total mass or the mass enclosed by the NM photosphere. But such confusion would not affect our results because the anomalous WDs we are considering is far from the Chandrasekhar limit.} The nature of DM admixed in these peculiar WDs could then be constrained by measuring their radii and rotational periods.


\subsection{SGRs and AXPs} \label{subsec:xray}

\citet{2013ApJ...762..117B} pointed out that massive, fast (rigidly) rotating, and highly magnetized WDs are candidates for soft gamma-ray repeaters (SGRs) and anomalous X-ray pulsars (AXPs) \citep{2012PASJ...64...56M}. These high-energy objects have rotation periods of $2 \lesssim P \lesssim 12$ s. However, the fact that an ordinary WD rotating with a period of $2$ s would be close to the Chandrasekhar limit casts doubts on whether these peculiar rotators could be stable-rotating, ordinary WDs. \\

\begin{table}
\caption[]{Critical rotation period in second for a rigidly-rotating DMRWD at the central density of core ignition (SNe, $\rho_{\text{SNe}} = 2 \times 10^{9}$ g cm$^{-3}$), core electron capture (EC, $\rho_{\text{EC}} = 10^{10}$ g cm$^{-3}$), and inverse beta decay (IBD, $\rho_{\beta} = 1.37 \times 10^{11}$ g cm$^{-3}$). \label{tab:minperiod}}
$$ 
\begin{array}{cccc}
\hline
\noalign{\smallskip}
\epsilon & P_{\text{Crit}}^{\text{SNe}} & P_{\text{Crit}}^{\text{EC}} & P_{\text{Crit}}^{\text{$\beta$}} \\
\noalign{\smallskip}
\hline
\noalign{\smallskip}
0.00 & 2.307 & 1.175 & 0.361 \\
0.05 & 2.303 & 1.142 & 0.341 \\
0.1 & 2.181 & 1.070 & 0.324\\
0.2 & 1.916 & 0.943 & 0.303 \\
\noalign{\smallskip}
\hline
\end{array}
$$ 
\end{table}

\begin{figure} 
	\centering
	\includegraphics[width=1.0\linewidth]{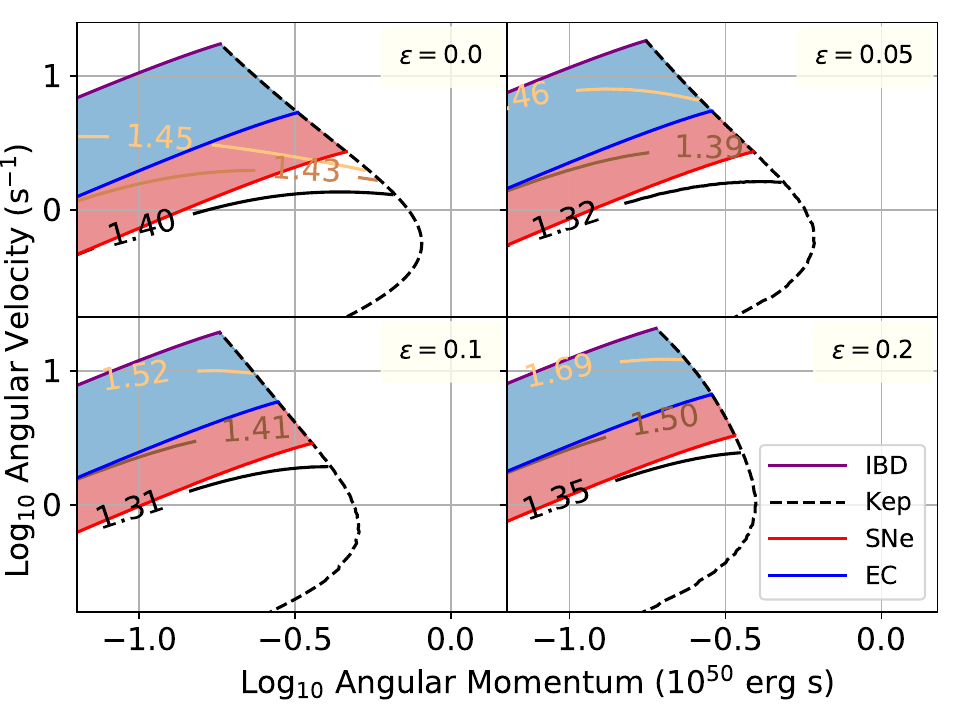}
	\caption{$\omega - J$ (both in log$_{10}$ scale) relations for DMRWDs with different DM mass fractions $\epsilon$. The red, blue, and purple lines mark the boundaries for DMRWDs reaching a central density of $\rho_{\text{SNe}}$, $\rho_{\text{EC}}$, and $\rho_{\beta}$, respectively, where $\rho_{\text{SNe}}$, $\rho_{\text{EC}}$, and $\rho_{\beta}$ are the central densities for core ignition, core electron-capture, and inverse beta-decay, respectively. The red (blue) shaded region indicates where DMRWDs are likely to exhibit core ignition (core electron capture). The black dashed line represents rigidly-rotating models at their Keplerian limit. We also plot the iso-mass contour lines, and their values represent the critical masses for SNe, EC, and IBD (listed in Table \ref{tab:criticalmass}). \label{fig:omegaj}}
\end{figure}

\begin{figure} 
	\centering
	\includegraphics[width=1.0\linewidth]{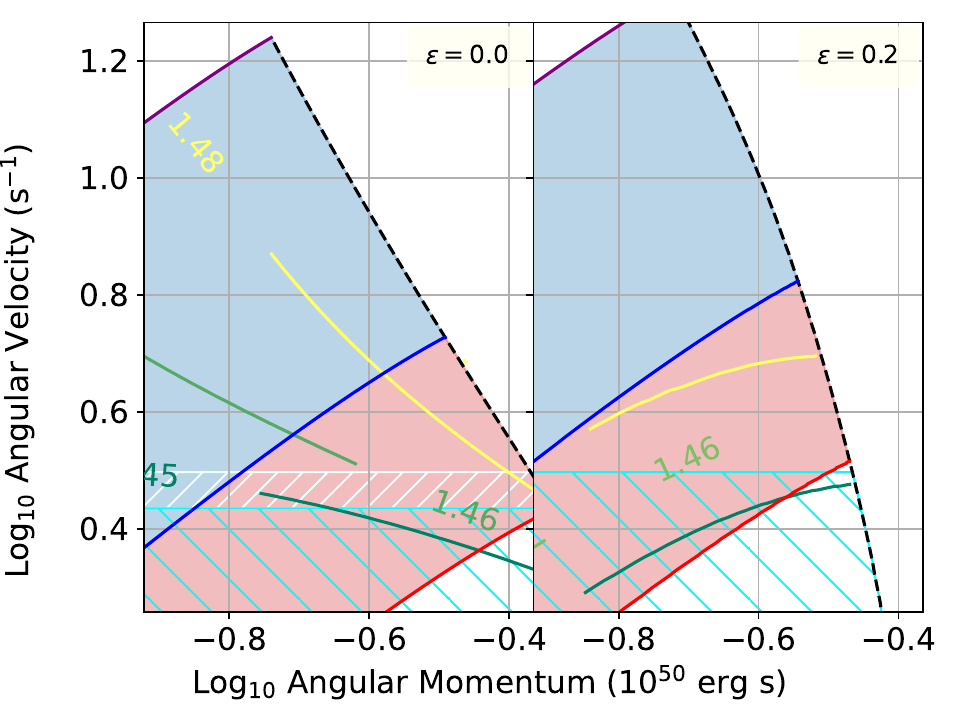}
	\caption{Same as Figure \ref{fig:omegaj}, but for WD (left) and DMRWDs with $\epsilon = 0.2$ (right) only. In the left panel, the white (aqua) band represents DMRWD models that rotate with a period $2 < P < P_{\text{Crit}}^{\text{SNe}}$ ($P > P_{\text{Crit}}^{\text{SNe}}$). In the right panel, the aqua band represents DMRWD models that rotate with a period $> 2$ s. \label{fig:fastwd}}
\end{figure}

To further investigate this issue, we follow \citet{2013ApJ...762..117B} to compute the $\omega - J$ relations for a pure NM, rigidly-rotating WD and show them in the upper-left panel of Figure \ref{fig:omegaj}. In the same figure, we show constant density boundaries for catastrophic events that a WD could experience: 1. core ignition (SNe, $\rho_{\text{SNe}} = 2 \times 10^{9}$ g cm$^{-3}$), 2. core electron capture (EC, $\rho_{\text{EC}} = 10^{10}$ g cm$^{-3}$), and 3. inverse beta decay (IBD, $\rho_{\beta} = 1.37 \times 10^{11}$ g cm$^{-3}$ for cold Fermi gas EOS), beyond which no stable WD would exist \citep{2013ApJ...762..117B, 1983bhwd.book.....S}\footnote{We note that $\rho_{\text{SNe}}$, $\rho_{\text{EC}}$, and $\rho_{\beta}$ might vary according to the micro-physical model assumed. Nonetheless, we quote their canonical values to understand the evolution of massive WDs qualitatively.}. Fast rotators with $\omega > 1$ s$^{-1}$ are close to the SNe boundaries, which is consistent with our expectations. We define the critical rotation periods $P_{\rm Crit}$ for SNe, EC, and IBD as the Keplerian rotation period for a rigidly-rotating WD right at the corresponding central density. We show these values for the pure NM model in the first column of Table \ref{tab:minperiod}. We observe that $P_{\text{Crit}}^{\text{SNe}} = 2.307$ s. Therefore, any SGRs or AXPs that are deemed to be consistent with a pure NM WD and with $2 \lesssim P \lesssim 2.307$ s would be near- to super-Chandrasekhar WDs ($\sim 1.45 - 1.48$ $M_{\odot}$) lying in the core ignition region. This is shown in the left panel of Figure \ref{fig:fastwd} as a white band. \\

\begin{figure} 
	\centering
	\includegraphics[width=1.0\linewidth]{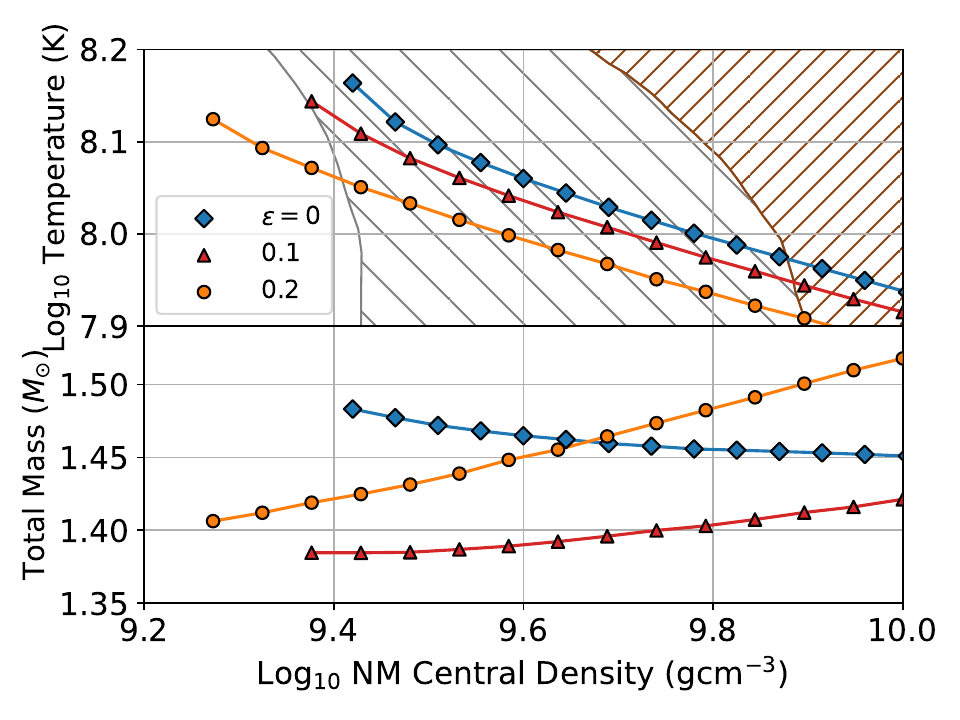}
	\caption{The central temperature is estimated by using Equation \ref{eqn:tempc} (upper panel) and total mass (lower panel) against central density for some DMRWDs rigidly rotating at $P = 2.07$ s, with DM mass fractions $\epsilon = 0$ (blue), $0.1$ (red), and $0.2$ (orange). We assumed a surface temperature of $10^{5}$ K. The grey (brown) boundary is the carbon ignition (carbon-oxygen fusion) curve taken from \citet{1984ApJ...286..644N} (\citet{PhysRevC.74.035803}). The hatched area represents the forbidden region for DMRWDs to exist stably without exploding as supernovae. Note that the grey-hatched area extends beyond the brown boundary. \label{fig:tempc}}
\end{figure}

Such a result poses two potential problems in explaining SGRs/AXPs using the WD model. First, a pure NM WD rotates at $2 \lesssim P \lesssim 2.307$ would lie in the core ignition region. Thus, for the pure NM WD to avoid core ignition, it should have low density and temperature. In particular, we notice one example of SGR/AXP from \citet{2012PASJ...64...56M} named 1E 1547-54 that might be inconsistent with being an ordinary, rigidly-rotating WD. This object has a rotating period of $2.07$ s. Assuming 1E 1547-54 to be a pure NM, rigidly rotating WD, could it survive exploding as a Type Ia supernova? We estimate the central temperature $T_{c}$ of our rigidly rotating DMRWDs by \citep{1976A&A....52..415K, 2021ApJ...921..138N}:
\begin{equation} \label{eqn:tempc}
    \frac{T_{\text{eff}}^{4}}{g} = 2.05 \times 10^{-10}T_{c}^{2.48}.
\end{equation}
Here, $T_{\text{eff}}$ is the effective surface temperature\footnote{The surface temperature of 1E 1547-54 is still unknown, and therefore our discussion only serves as a rough estimation.}, which we set to $10^{5}$ K, and $g$ is the surface gravity. We show the estimation in Figure \ref{fig:tempc}. In the same figure, we append the carbon ignition curve (grey) and carbon-oxygen fusion curve (brown), taken from \citet{1984ApJ...286..644N} and \citet{PhysRevC.74.035803}, respectively. It is believed that on the right of the grey line, carbon ignition would lead to a thermonuclear runaway. We observe that all massive super-Chandrasekhar models are too hot to exist stably as a rotating WD. \\

The understanding of thermonuclear supernova progenitors and explosions is still highly uncertain. Given the large diversity of Type Ia supernova light curves, it is natural to expect a large variety of supernova progenitors and hence a large range of central $\rho$ and $T$. In other words, the carbon ignition curve might shift. In addition, pycnonuclear burning occurs at a long time scale, so less massive models with $M \sim 1.45$ $M_{\odot}$ might be able to escape from being a supernova. However, the second issue is that such a value might be too large. The most massive WD discovered so far has a mass of $1.365$ $M_{\odot}$ \citep{Caiazzo2021}. As we discussed earlier, super-Chandrasekhar WDs with mass $\gtrsim 1.46$ $M_{\odot}$ discovered in \citet{https://doi.org/10.48550/arxiv.1610.00986} are highly suspicious due to the lack of error bars. Therefore, the expected mass of a WD being a rapidly-rotating AXP/SGR would challenge the well-established data. Theoretically, it is also difficult to form a WD with $M \sim 1.45$ $M_{\odot}$. For instance, by running accreting Carbon WD models with mass accretion rates from $10^{-7}$ to $10^{-11}$ $M_{\odot}$/year using the MESA stellar evolution code \citep{Paxton2011, Paxton2013, Paxton2015, Paxton2018, Paxton2019}, we found that the maximum mass of an igniting, pure NM WD is $M \sim 1.4$ $M_{\odot}$. Therefore, even if 1E 1547-54 is a $M \sim 1.45$ $M_{\odot}$ WD, it is unlikely to be formed by mass accretion. It is also unlikely that 1E 1547-54 is a super-Chandrasekhar WD formed by a binary WD merger because \citet{2016MNRAS.463.3461S} showed that massive, super-Chandrasekhar WD mergers would end up as either 1. supernovae, 2. sub-Chandrasekhar WDs, or 3. neutron stars. Therefore, 1E 1547-54 should not be a normal WD. \\

Although a WD model is preferred \citep{2012PASJ...64...56M}, the true identity of 1E 1547-54 is still unknown. We show that this object might be a rigidly-rotating DMRWD. To confirm our hypothesis, we compute the $\omega - J$ relations for rigidly-rotating DMRWD with different $\epsilon$ and show them in Figure \ref{fig:omegaj}. We also compute their critical rotation periods and list them in Table \ref{tab:minperiod}. We observe that the constant density boundaries for catastrophic events (SNe, EC, and IBD) shift upward when $\epsilon$ increases - admixing DM makes the NM component at a fixed $\rho_{c}$ rotate faster. Thus, $P_{\text{Crit}}^{\text{SNe}}$, $P_{\text{Crit}}^{\text{EC}}$, and $P_{\text{Crit}}^{\beta}$ (c.f. Table \ref{tab:minperiod} for their definitions) all decrease as $\epsilon$ increases, allowing for more DMRWDs to rotate faster without reaching $\rho_{\text{SNe}}$. We show in the right panel of Figure \ref{fig:fastwd} that if we assumed $\epsilon = 0.2$, then a DMRWD rotating with $P = 2.07$ s is indeed possible. We use Equation \ref{eqn:tempc} to estimate the core temperature of these models and find that they lie outside the forbidden region in Figure \ref{fig:tempc}. These models are also less massive ($\sim 1.4$ $M_{\odot}$). We remark that the effective temperature of pulsars spans from $10^{5}$ to $10^{7}$ K \citep{2011AIPC.1379...60N}. We find that the $T_{c}$ reaches $10^{9}$ K if $T_{\rm eff} \sim 10^{5.5}$ K. Carbon ignition would initiate at such a high temperature. Therefore, 1E 1547-54 could be a rigidly-rotating DMRWD if its $T_{\rm eff}$ is $10^{5}$ K or less, which is on the lower end of the typical range of pulsar surface temperature. \\

The observed spinning down of 1E 1547-54 provides additional evidence for its non-ordinary structure. As shown by \citet{2013ApJ...762..117B} that under the constant-mass assumption, some super-Chandrasekhar WDs would spin up (increase in $\omega$) as they lose $J$. Such a phenomenon has been observed by \citet{1990ApJ...357L..17S} and \citet{2000ApJ...534..359G}. Here, we reproduce their results in Figure \ref{fig:fastwd}. Following the iso-mass contours on the figure, we observe a handful of super-Chandrasekhar WDs spinning up as they gradually lose $J$. These models are all super-Chandrasekhar WDs so massive that degenerate pressure alone cannot support their structures. So they rotate faster to provide stronger centrifugal force as they lose $J$. However, several follow-up studies on 1E 1547-54 have revealed that it is spinning down \citep{2016arXiv160300141M}\footnote{See also \url{http://www.physics.mcgill.ca/~pulsar/magnetar/main.html}}, which is inconsistent with the fact that pure NM WDs with a period of $2.07$ s should be spinning up as it loses angular momentum (c.f. left panel of Figure \ref{fig:fastwd}). This also casts doubts on whether 1E 1547-54 could be an ordinary, rigidly-rotating WD. We show that for DMRWD, there are fewer super-Chandrasekhar WDs that would spin up. For instance, all models with $\epsilon = 0.2$ will only spin down as they lose $J$ (c.f. Figure \ref{fig:omegaj}). This is because admixing DM would lift the non-rotating Chandrasekhar limit (c.f. Figure \ref{fig:database}). Since DMRWD rotating at a period of $2.07$ is spinning down, this suggests that 1E 1547-54 could be a DMRWD. Nonetheless, 1E 1547-54 could be an example of anomalously rapid-rotating WDs, and we believe such objects deserve more follow-up analysis to reveal their identities. We should remark that WDs have been observed as pulsars \citep{2016ApJ...831L..10G, Buckley2017}, and the possibility that 1E 1547-54 is a WD-like object should not be ruled out. \\

As we discussed earlier, understanding the structure of rotating WDs would be essential to predicting their final fates qualitatively \citep{2005A&A...435..967Y}. Because some studies \citep{Leung_2015, Chan_2021} used the DM-admixed Type Ia supernova model to account for peculiar transients, it would be interesting to see if admixing DM would alter the evolution path of WD toward thermonuclear runaway. We, therefore, presented some preliminary analysis in Appendix \ref{sec:barrier}. Furthermore, it is important to know whether the DMRWD we considered here is stable against secular instability. We have examined the stability of DMRWDs, and they are presented in Appendix \ref{sec:stab}.


\subsection{Applications of the $I-$Love$-Q$ Relations} \label{subsec:iloveq}
This section focuses on the effects of $\epsilon > 0$ on the well-known $I-$Love$-Q$ universal relations for WDs.

\subsubsection{The Deviation of the Universal Relations} \label{subsec:deviate}
We have computed the $I-$Love$-Q$ relations for rigidly-rotating DMRWDs using several EOSs mentioned in section \ref{sec:eos}. We construct a few sequences of DMRWD models by varying the maximum NM density $\rho_{\text{Max}2}$ from $10^{6}$ g cm$^{-3}$ to $\sim$ $10^{9}$ g cm$^{-3}$ for fixed $\epsilon$ ranging from $0$ to $0.3$. \\

\begin{figure*}[htb!]
\gridline{\fig{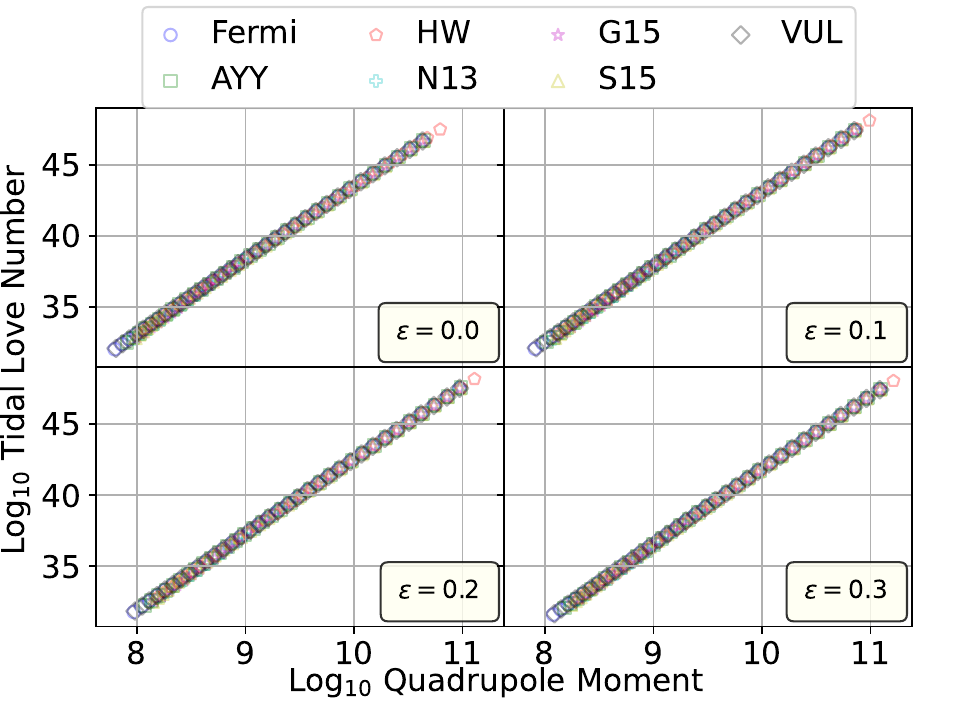}{0.5\textwidth}{(a)}
          \fig{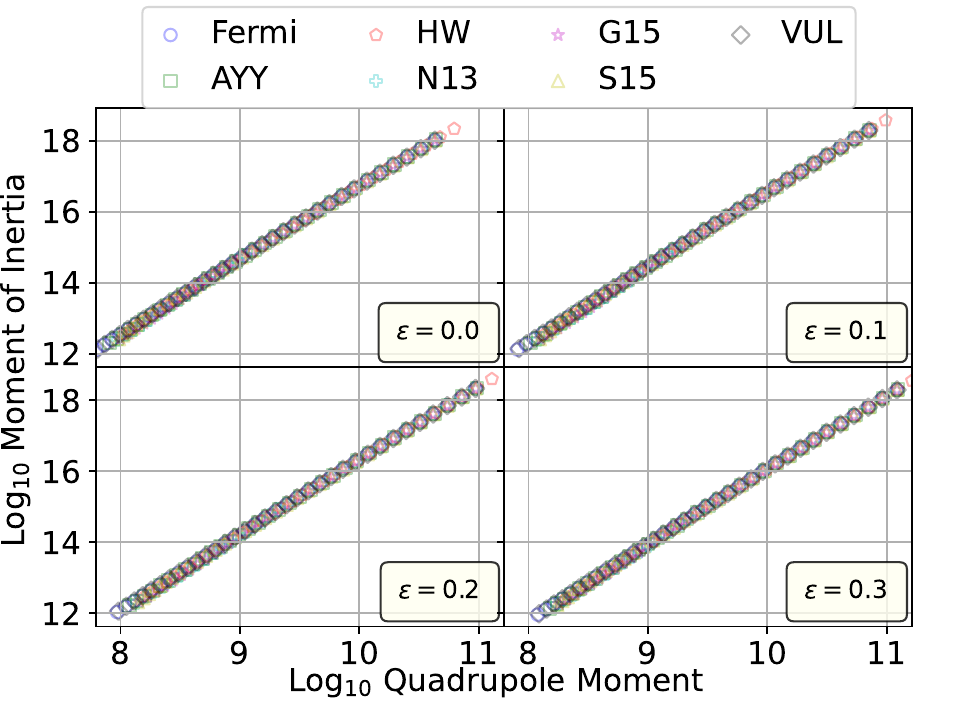}{0.5\textwidth}{(b)}
          }
\caption{(a) Scatter plots of log$_{10}\lambda_{T}$ against log$_{10}Q$ for slowly, rigidly rotating DMRWDs with different EOS assumed. We included 4 subplots to show that the EOS-independent relation remains valid when DM is admixed. The lower right legend indicates the mass fraction $\epsilon$ of DM admixture. (b) Same as (a), but for log$_{10}I$ against log$_{10}Q$. The legend for both plots (i.e., Fermi, HW, AYY, etc ...) is the EOS assumed (c.f. Section \ref{sec:eos}.). \label{fig:iloveqscat}}
\end{figure*}

\begin{figure*}
\gridline{\fig{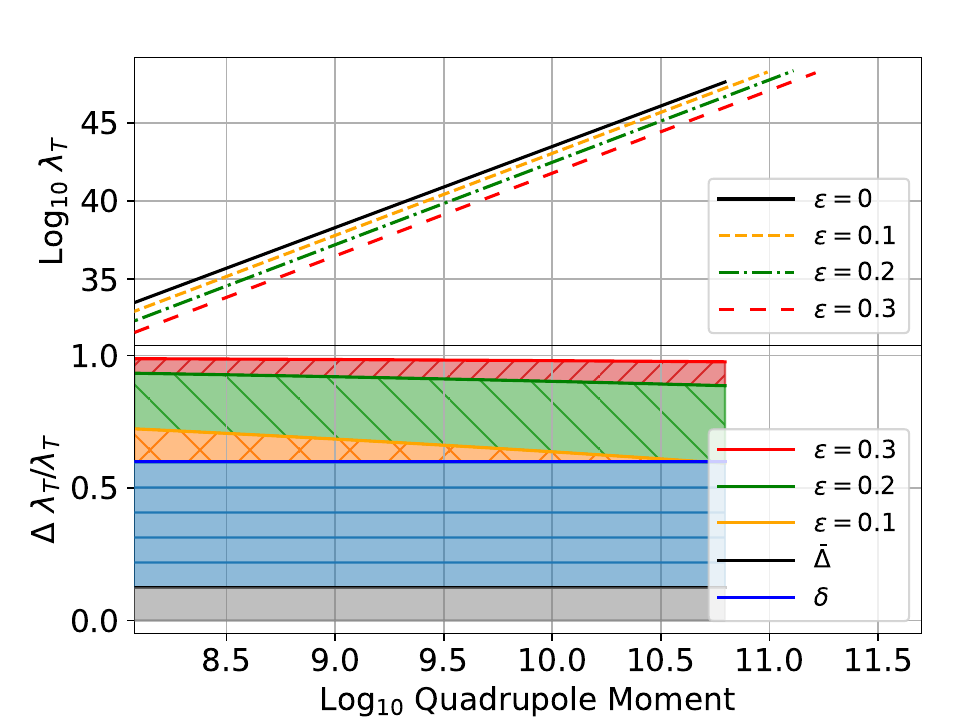}{0.5\textwidth}{(a)}
          \fig{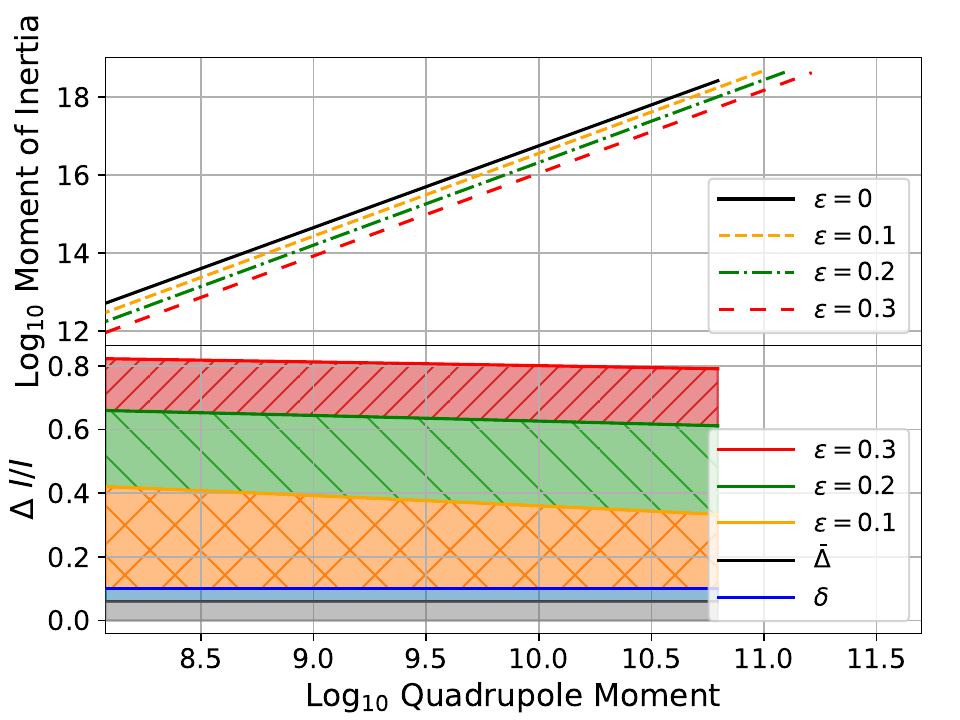}{0.5\textwidth}{(b)}
          }
\caption{(a) \textit{Upper panel}: Best-fit linear lines of log$_{10}\lambda_{T}$ vs. log$_{10}Q$ relation for slowly, rigidly rotating DMRWDs having different DM mass fractions $\epsilon$. The best-fit linear line is generated by fitting scatter points in Figure \ref{fig:iloveqscat}. \textit{Lower panel}: $\Delta\lambda_{T}/\lambda_{T}$ = ($\lambda_{T}^{\text{DM}}$ - $\lambda_{T}^{\text{NM}}$)/($\lambda_{T}^{\text{NM}}$) vs. $\log_{10}Q$ for different $\epsilon$. $\Delta\lambda_{T}$ is computed by taking the difference between the best-fit line of a DM-admixed model with the pure NM model. The solid grey line indicates the average fitting uncertainty $\bar{\Delta} = 0.16$ of the $\text{Love} - Q$ relations for the pure NM model. The blue solid line represents estimated measurement uncertainty $\delta = 0.6$ as taken from \citet{2013PhRvD..88b3009Y} and \citet{2013LRR....16....9Y}. We fill the area between solid lines to form colored strips. We interpret this figure in such a way: for instance, WDs lying in the green strip have $0.1 < \epsilon < 0.2$, and WDs lying in the blue (grey) strip are within the measurement (fitting) uncertainty. (b) Same as (a), but for log$_{10}I$ vs. log$_{10}Q$, with $\bar{\Delta} = 0.08$ and $\delta = 0.1$. \label{fig:iloveqrigid}}
\end{figure*}

We show the scatter plot obtained by assuming different EOS in Figure \ref{fig:iloveqscat}. In particular, the upper-left panel of Figure \ref{fig:iloveqscat} (a) and (b) shows that, when there is no DM admixture, the $I-$Love$-Q$ relations for WDs computed by using different EOSs lie approximately on the same line. In particular, we find the maximum deviations from the best-fit line are within $\sim 1$ \% in the log$_{10}$ scale. These results also agree with previous studies on the $I-$Love$-Q$ relations of slowly rotating WDs, particularly for the universality lines and the corresponding maximum deviations. We also observe a small mass fraction (e.g., $\epsilon \sim 0.01$, not shown in Figure \ref{fig:iloveqscat}) of DM admixture makes no change to the $I-$Love$-Q$ universality. Most of the deviations due to the DM admixture are within the average fitting uncertainties of the pure NM model. However, we observe two features when $\epsilon$ increases beyond $0.01$. First, Figure \ref{fig:iloveqscat} shows that the $I-$Love$-Q$ relations remain independent of EOSs assumed. Second, the universal lines are all shifted to higher $Q$ values relative to those with $\epsilon = 0$, which is shown in the upper panels of Figure \ref{fig:iloveqrigid} (a) and (b). The relative deviations between the universal Love$-Q$ ($I-Q$) relations for DMRWDs and those for pure NM models are getting larger than the corresponding average fitting uncertainty $\bar{\Delta} = 0.16$ ($0.08$) of the latter, and the deviation increases with $\epsilon$. We fit the $I-$Love$-Q$ relations for DMRWD with different $\epsilon$ using a log-linear function:
\begin{equation}
    \text{log}_{10}y = a + b\text{log}_{10}x,
\end{equation}

\begin{figure}
	\centering
	\includegraphics[width=1.0\linewidth]{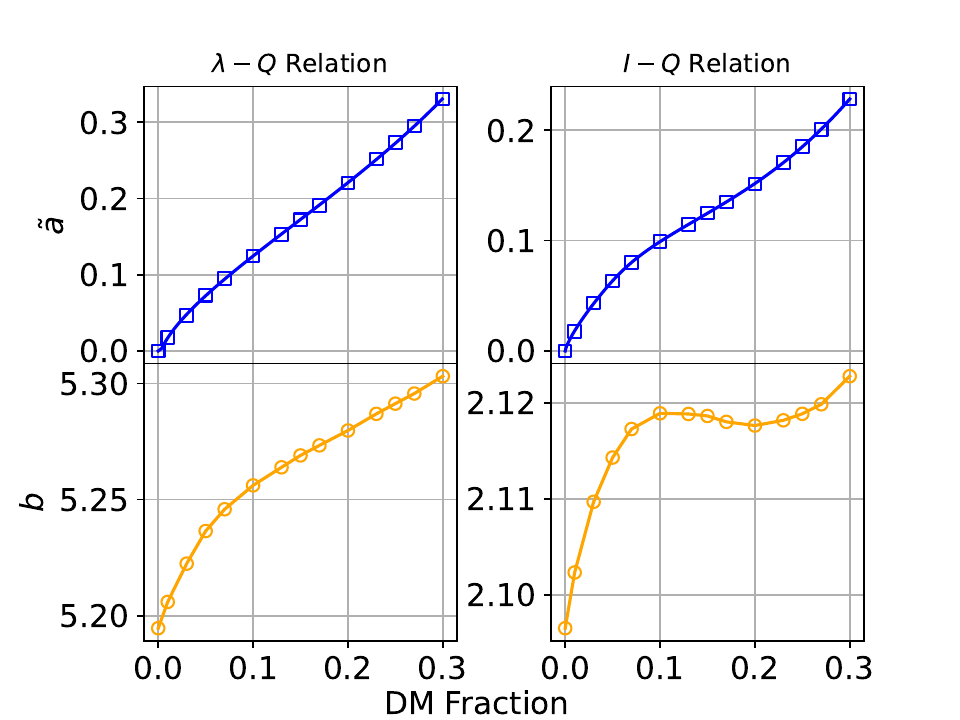}
	\caption{Correlation between parameters $\tilde{a} = (a' - a)/a$ and $b$ with the DM mass fraction $\epsilon$ for the $\lambda - Q$ relation (left panel) and the $I - Q$ relation (right panel). Blue lines are fitting functions presented in Equation \ref{eqn:fit}, while orange lines are just discrete linear lines connecting the data points. \label{fig:abparameter}}
\end{figure}

\begin{table}
\caption[]{Parameters $a$ and $b$ for fitting the universal $\lambda - Q$ and $I - Q$ relations of DMRWDs for different $\epsilon$. \label{tab:abpara}}
$$ 
\begin{array}{ccccc}
\hline
\noalign{\smallskip}
\epsilon & a_{\lambda - Q} & b_{\lambda - Q} & a_{I - Q} & b_{I - Q} \\
\noalign{\smallskip}
\hline
\noalign{\smallskip}
0.00 & -8.77 & 5.23 & -4.36 & 2.11 \\
0.01 & -8.95 & 5.24 & -4.45 & 2.12 \\
0.05 & -9.48 & 5.28 & -4.65 & 2.13 \\
0.1 & -9.90 & 5.30 & -4.78 & 2.13 \\ 
0.2 & -10.67 & 5.32 & -4.97 & 2.13 \\
0.3 & -11.54 & 5.34 & -5.29 & 2.13 \\
\noalign{\smallskip}
\hline
\end{array}
$$ 
\end{table}

with $y$ being $\lambda_{T}$ or $I$ and $x = Q$. We show the results in Figure \ref{fig:abparameter} and Table \ref{tab:abpara}. The parameter $a$ is strongly correlated with $\epsilon$, while $b$ shows a much weaker correlation, reflecting the negligible changes in the slopes of the $I-$Love$-Q$ relations for different $\epsilon$.

\subsubsection{Detection Prospect} \label{subsec:detect}
Gravitational waves are emitted in binary WD merger events \citep{2014ApJ...794...35G, 2020MNRAS.491.3000M, 2021ApJ...906...29Y}. Although the signal is below detection thresholds in the current gravitational-wave detectors, it was estimated that a binary WD merger event could be resolvable for the next generation of detectors \citep{korol, yu}. In fact, a method for estimating WD masses by extracting the tidal information from the gravitational-wave signatures has been proposed \citep{10.1093/mnrasl/slaa183}. Hence, we anticipate that measurements of the $I-$Love$-Q$ relations for WDs could help reveal the existence of DM in rotating WDs. \\

In reality, each WD may acquire a different amount of DM. Therefore, WDs would scatter within a band of $I-$Love$-Q$ relations spanned by different $\epsilon$. We show these bands together with the uncertainties of the pure NM models in the lower panels of Figure \ref{fig:iloveqrigid} (a) and \ref{fig:iloveqrigid} (b). They are computed by first taking differences between the best-fit lines for DM-admixed models with those of the pure NM models, then plotting the differences versus $Q$ as solid lines. Finally, we fill in the area between solid lines to form different bands. In particular, since the uncertainties (shown as the grey band) of the pure NM models are small, any significant deviation that lies above the pure NM version of the $I-$Love$-Q$ relations is possibly a sign of a sub-solar mass scale of DM admixture. \\

Are DMRWDs detectable by measuring the deviation from the $I-$Love$-Q$ relations? As mentioned in \citet{2013PhRvD..88b3009Y} and \citet{2013LRR....16....9Y}, $\lambda_{T}$ and $I$ could be measured by gravitational-wave and/or electromagnetic observations. We take their estimated uncertainties as $60$ \% and $10$ \%, respectively. We compare these values to the deviations in $\lambda_{T}$ and $I$ in the lower panels of Figure \ref{fig:iloveqrigid} (a) and \ref{fig:iloveqrigid} (b), respectively. We find it possible to detect DMRWDs with $\epsilon \geq 0.1$. Future space-based detectors should have improved accuracy to detect or constrain DMRWDs with $\epsilon < 0.1$. \\

Suppose there is an anomalous detection of $\lambda_{T}$ or $I$ with respect to a fixed $Q$. If such a deviation is a result of DM admixture, we can infer the value of $a$ from $\Lambda_T$ or $I$:
\begin{equation} \label{eqn:change}
    a' \approx \text{log}_{10} x' - b_{0}\text{log}_{10} Q'.
\end{equation}
Here, the prime variables refer to anomalous detection, and those with subscript zero the pure NM model with $x = \lambda_{T}$ or $I$. We also assumed that the best-fit parameter $b$ of the DM-admixed models does not vary significantly from that of the pure NM model, so we can substitute their values with that of the pure NM model. We fit the relation between $a$ and $\epsilon$ for the $\lambda_{T}-Q$ and $I-Q$ universal curve and obtain:
\begin{equation} \label{eqn:fit}
\begin{aligned}
   \epsilon(\tilde{a}_{\lambda_{T} - Q}) = 0.468\text{tanh}(2.756\tilde{a}_{\lambda_{T} - Q})^{1.461} \\ - 0.012\text{tanh}(\tilde{a}_{\lambda_{T} - Q})^{0.239}, \\
   \epsilon(\tilde{a}_{I - Q}) = 0.433\text{tanh}(2.984\tilde{a}_{I - Q})^{1.296} \\ + 0.084\text{tanh}(14.832\tilde{a}_{I - Q})^{16.831}.   
\end{aligned}
\end{equation}
For $0 \leq \epsilon \leq 0.3$. Here, $\tilde{a} = (a' - a)/a$ is the relative change in $a$ with respect to the pure NM model. Using the above equation, one can directly infer the mass fractions $\epsilon$ of DM admixture in a rotating WD.

\subsubsection{Concluding Remark} \label{subsec:remark}
\citet{2020MNRAS.492..978T} demonstrated that the $I-$Love$-Q$ relations for white dwarfs depend on the degree of differential rotation (i.e., the angular velocity profile). However, we have computed the $I-$Love$-Q$ relation for DMRWDs, assuming different rotation profiles. They include the $j$-const, $v$-const, and ``Kepler'' profiles (c.f. \citet{1986ApJS...61..479H} and \citet{Yoshida_2019}). We still observe that the $I-$Love$-Q$ relations for DMRWD are universal but differ from those of the pure NM, rotating WD. Assuming that the rotation rules that we assumed are representative enough, we conclude that the deviations of the DM-admixed $I-$Love$-Q$ relations from those of the pure NM model are insensitive to whether the WD is rotating rigidly or differentially. We note that the universality of the $I-$Love$-Q$ relations is violated for hot WDs \citep{10.1093/mnras/sty1227}. However, our DMRWD models produce deviations from the NM version of the universal relations that are rather different from those of hot WDs.

\subsection{Heavier/lighter Dark Matter Particles?} \label{subsec:heavylight}

\citet{Chan_2021} and \citet{PhysRevD.105.123010} showed that the properties of Fermionic DM-admixed compact stars change sharply around a DM particle mass of $0.1$ GeV. Therefore, our results should be sensitive to the choice of such value. We reproduce the analysis presented in Sections \ref{subsec:xray} and \ref{subsec:iloveq} and give qualitative descriptions of how our results would change according to the DM particle mass. In particular, we choose 0.2 GeV (0.08 GeV) to represent the heavy (light) DM limit. \\ 

If we assume a particle mass of 0.2 GeV, then the DM acts as a compact core. We find the WD total mass and critical rotation velocity to be reduced. So there will be fewer DMRWDs that can rotate at 2.307 s and be free from the thermonuclear runaway. We also find the $I$-Love-$Q$ relations to deviate more than that of the 0.1 GeV case. Also, the $I$-Love-$Q$ relations break down when the log quadrupole moment is small; The results for assuming a particle mass of 0.08 GeV are vice versa - the DM component usually extends to a larger radius than the NM component. We find the critical rotation velocity increased. So, there are more available DMRWDs that can rotate at a period of 2.307 s and be free from the thermonuclear runaway. However, these models have a huge DM mass. It is doubtful whether such massive WDs exist. Also, the $I$-Love-$Q$ relations are still universal, showing few differences from the 0.1 GeV case. However, when the DM component is much larger than the NM, it would be difficult to correctly reproduce the composite $I$/Love/$Q$ numbers.


\section{Conclusion} \label{sec:conclusion}
Earlier studies on DM-admixed neutron stars and supernovae could explain a handful of peculiar astrophysical objects. In this paper, we investigate the observational signatures of DMRWDs. \\

We first compute the mass-radius relations for rigidly-rotating DMRWDs. Our DMRWD models could successfully account for some anomalous low, or high-mass WDs observed. We then show that admixing DM increases the rotational velocity of a WD, thus allowing the existence of some stable, rapidly-rotating WD that could be free from the thermonuclear runaway. These DMRWDs have a near-Chandrasekhar mass and are spinning down, which is consistent with some fast-rotating SGRs/AXPs (such as 1E 1547-54). We further discover that DMRWDs follow universal $I-\text{Love}-Q$ relations that deviate significantly from those of the pure NM models, with a larger deviation for a larger $\epsilon$. Since each WD may have a different DM fraction, the WD $I-\text{Love}-Q$ relations span bands above the universal lines for the pure NM models, and the deviations for $\epsilon \ge 0.1$ are large enough to be observable. Finally, we give an empirical formula to extract the DM mass fractions in DMRWDs from the deviations of the $I-\text{Love}-Q$ relations with respect to those of the pure NM model. \\

Several issues remain to be resolved. First, we have assumed that the DM is non-rotating. Even though accretion can change the collective motion of the DM component, it was pointed out by \citet{Iorio_2010} that a neutron star in the galaxy could accrete DM at a rate of $10^{7}$ kg s$^{-1}$. This value is so small that there could be no considerable amount of DM build-up at the outer envelope in the time scale of the age of the Universe. Thus, the non-rotation approximation shall remain valid throughout the evolution from the zero-age main sequence to the formation of DMRWD. However, we remark that degenerate pressure is an effective quantum self-interaction. Thus, the DM component may have inherited collective motion during the molecular cloud collapse. The gravitational force from the NM component could also drag the DM component. Nonetheless, we assume the DM component to be non-rotating for this exploratory study. \\

Second, our analysis is solely based on the Newtonian framework. As the compactness of WDs continues to increase when one considers $\rho \rightarrow 3\times10^{10}$ g cm$^{-3}$, general relativistic effects would become more important. We expect our results computed beyond such a value to alter when general relativity is considered. Still, results computed around $\rho \sim \rho_{\text{SNe}}$ shall be mildly affected. To extend our work to the general relativistic regime, one needs to include a second component of non-self-interacting DM into the matter source term of the Hartle–Thorne metric \citep{1967ApJ...150.1005H}, or one can solve a set of self-consistent two-fluid elliptic equations to compute the structure of relativistic DMRWDs. These could also be used to examine DM-admixed rotating neutron stars. \\

In conclusion, we predict the observational signatures of DM admixtures in rotating WDs and connect our results to some peculiar compact objects. Our results could be applied to study the potentially large number of unusual compact objects that could be discovered by next-generation, LSST-like surveys.


\begin{acknowledgements}
We thank stellarcollapse.org for providing the electron fraction table for the VUL EOS. We acknowledge Dr. Lap-Ming Lin for providing valuable comments and suggestions on the $I-$Love$-Q$ relations and for choosing WD EOS. This work is partially supported by a grant from the Research Grant Council of the Hong Kong Special Administrative Region, China (Project No. 14300320). Shing-Chi Leung acknowledges support from NASA grants HST-AR-15021.001-A and 80NSSC18K1017.
\end{acknowledgements}

\appendix


\section{Connections to the DM-admixed Supernova Model} \label{sec:barrier}

\begin{table}
\caption[]{Same as Table \ref{tab:minperiod}, but for the critical total mass (all in units of $M_{\odot}$). The critical mass for SNe, EC, and IBD is defined as the mass of a non-rotating DMWD at the corresponding density. See Table \ref{tab:minperiod} for the definition of these densities.} \label{tab:criticalmass}
$$ 
\begin{array}{cccc}
\hline
\noalign{\smallskip}
\epsilon & M_{\text{Crit}}^{\text{SNe}} & M_{\text{Crit}}^{\text{EC}} & M_{\text{Crit}}^{\text{$\beta$}} \\
\noalign{\smallskip}
\hline
\noalign{\smallskip}
0.00 & 1.40 & 1.43 & 1.45 \\
0.05 & 1.32 & 1.39 & 1.46 \\
0.1 & 1.31 & 1.41 & 1.52 \\
0.2 & 1.35 & 1.50 & 1.69 \\
\noalign{\smallskip}
\hline
\end{array}
$$ 
\end{table}

\begin{figure} 
	\centering
	\includegraphics[width=1.0\linewidth]{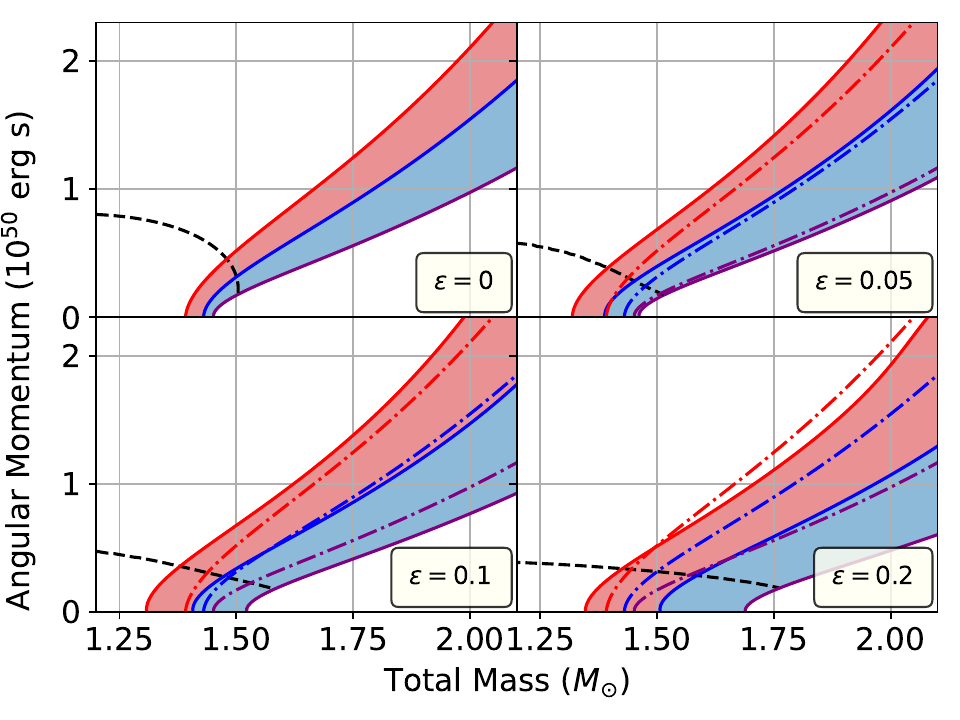}
	\caption{Same as Figure \ref{fig:omegaj}, but for the angular momenta against total masses. We include rigid and differential rotation models in constructing the colored boundaries. Dash-dotted lines represent boundaries computed for the pure NM model. \label{fig:mjplot}}
\end{figure}

\citet{2005A&A...435..967Y} analyzed pre-supernova evolution by considering the $J - M$ trajectory. In their study, a series of constant NM central density models are fitted and plotted in the $J - M$ diagram. For a given $M$ of the accreting WD, the angular momentum threshold for forming a thermonuclear explosion would be the amount of $J$ to lose so that a WD would end up having $\rho \sim \rho_{\text{SNe}}$. Furthermore, since carbon ignition would be initiated around temperature $T$ of $\sim 10^{8} - 10^{8.5}$ K at $\rho \sim \rho_{\text{SNe}}$, another criterion for an accreting WD to become a supernova is that the temperature can support nuclear runaway once it reaches $\rho \sim \rho_{\text{SNe}}$. This depends on the competition between the angular momentum loss time scale $\tau_{J}$ and the cooling time scale $\tau_{\text{cool}}$. Since $\tau_{J}$ must depend on the amount of $J$ to be lost, a smaller angular momentum barrier would mean a shorter $\tau_{J}$, and the WD could remain at a high $T$ once it reaches $\rho \sim \rho_{\text{SNe}}$. \\

Here, we extend the computation of rotating WDs to include an admixture of DM and show our results in Figure \ref{fig:mjplot}. To construct the full boundaries, we have included differential rotation profiles such as the $j$-const, $v$-const, and ``Kepler'' profile (c.f. \citet{1986ApJS...61..479H} and \citet{Yoshida_2019}). These differentially rotating profiles read:
\begin{equation}
    \omega^{2} \sim \frac{1}{(s^{2} + d^{2})^{\alpha}}.
\end{equation}
Here $\alpha$ varies according to different rules assumed. There is a free parameter $d$, which measures how differential the rotation is. We choose $d$ to be the numerical value of the equatorial radius where $\rho_{2} = 0.01$ that of the (NM) central value. It is chosen to match results by \citet{2005A&A...435..967Y}. When a small amount of DM admixture is present (say, $\epsilon \lesssim 0.1$), the angular momentum barrier (shown as the red line) for core ignition is shifted to a lower total mass. This leads to three consequences: 1. there is a larger range of $M$ available to initiate core ignition, and 2. DMRWDs need to lose less $J$ to reach $\rho \sim \rho_{\text{SNe}}$, and 3. it is easier for accreting WDs evolving from the low-mass region in the diagram to end up having $\rho \sim \rho_{\text{SNe}}$. \\

The shift of the angular momentum barrier to lower total mass is due to two reasons. First, the critical mass for core ignition $M_{\rm Crit}^{\rm SNe}$, defined as the mass of a non-rotating DMWD that has a central density of $\rho_{\text{SNe}}$, is reduced. We show $M_{\rm Crit}^{\rm SNe}$, $M_{\rm Crit}^{\rm EC}$, and $M_{\rm Crit}^{\beta}$ for different $\epsilon$ in Table \ref{tab:criticalmass}. Here, $M_{\rm Crit}^{\rm EC}$ is the critical mass for core electron capture, and $M_{\rm Crit}^{\beta}$ is the critical mass for inverse beta decay instability. Same as $M_{\rm Crit}^{\rm SNe}$, $M_{\rm Crit}^{\rm EC}$ and $M_{\rm Crit}^{\beta}$ are evaluated at $\rho = \rho_\text{EC}$ and $\rho_\beta$ respectively. We observe that $M{\text{Crit}}^{\text{SNe}}$ and $M_{\text{Crit}}^{\text{EC}}$ first decrease and then increase as $\epsilon$ is increasing. In contrast, $M_{\text{Crit}}^{\text{$\beta$}}$ is an increasing function of $\epsilon$. These trends have been observed by \citet{Chan_2021}. Whether admixing a small fraction of DM would lead to an increase or reduction of $M$ depends on the DM particle mass assumed. However, the increase of $M$ when $\epsilon$ is large could be understood by the following quantitative argument (in the Newtonian framework). At a fixed central density, and we vary the amount of DM admixture $M_{1}$ as a free parameter, $M$ would behave as:
\begin{equation}
    M = M_{2}(M_{1}) + M_{1}.
\end{equation} 
The hydrostatic equations implied that $M_{2}$ depends on $M_{1}$, which is a decreasing function bounded from below by $0$. When $M_{1}$ is large, $M_{2}$ will be close to $0$. As such, we have:
\begin{equation}
    M \approx M_{1}.
\end{equation} 
So $M$ scales linearly with $M_{1}$ for large $M_{1}$ in the Newtonian limit with no maximum $M_{1}$. This argument could also be applied to the case of admixing DM with a particle mass of over $1$ GeV, but the DM mass required to be admixed to increase the total mass will be large, and therefore these trends have not been observed by \citet{2013PhRvD..87l3506L}. \\

In addition, when $\epsilon$ is small, the $J$ for a given $M$ of DMRWD right at $\rho \sim \rho_{\text{SNe}}$ is larger than that of $\epsilon = 0$. Figure \ref{fig:profile} illustrates the DM and NM density profiles for different $\epsilon$. We observe that the DM component with a sub-GeV particle mass is more extended and diffused. Its central density is also lower than the NM by one order of magnitude. Therefore, a small amount of DM admixture would not significantly affect the structure of the NM component. For instance, the NM density profiles only change mildly with small $\epsilon$, except that the NM component rotates faster to balance the increase in gravitational attraction. However, as $\epsilon$ increases, the mass and radius of the NM component are reduced significantly, so that $J \sim \omega M_{2}R_{2}^2$ is reduced. \\

We observe that a small amount (say $\epsilon \lesssim 0.1$) of admixed DM favors the thermonuclear explosion scenario in two ways - it allows a larger range of masses and reduces the angular momentum barrier for a WD to reach the core ignition density $\rho_{\text{SNe}}$. These results might connect to the DM-admixed supernova models proposed by \citet{Leung_2015} and \citet{Chan_2021}. In particular, DM-admixed WDs with $0.1$ GeV DM particle mass were used by \citet{Chan_2021} as progenitors for studying supernova explosions. They showed that these exotic progenitors would produce dimmer and broader light curves compared with ordinary supernova models. This is because the photosphere evolves slower and loses thermal energy through pressure work done when expanding against an external DM potential. Their models match some peculiar supernovae with low-luminosity and slowly-evolving light curves, with a range of $\epsilon$ from $0.05$ to $\sim 0.33$. Other than stability, another challenging issue of having DM-admixed WDs as supernova progenitors is the likelihood of them exploding. This work shows that from the angular momentum perspective, DM-admixed WDs are more likely to explode than ordinary WDs, given that the DM admixture is not too large. While $\epsilon \approx 0.2$ is the marginal case, when $\epsilon$ reaches about $0.3$, such as in model DM3 presented by \citet{Chan_2021}, the angular momentum barrier becomes larger when compared with the pure NM model, which disfavors explosion.


\section{Stability Analysis} \label{sec:stab}

\begin{figure} 
	\centering
	\includegraphics[width=1.0\linewidth]{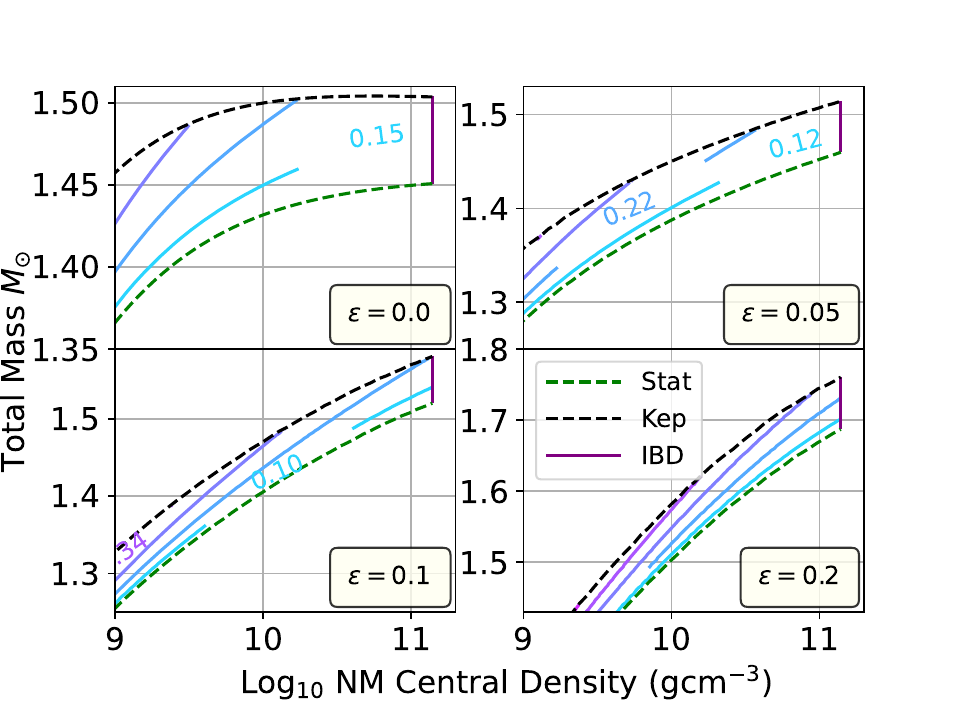}
	\caption{Total mass against central NM density for DMRWDs with different DM mass fractions $\epsilon$. The green lines are non-rotating DM-admixed models, while the black lines are rigidly rotating DMRWDs at their Keplerian limit. The purple lines mark the boundaries for inverse beta decay instabilities. We also plot the constant angular momentum lines to analyze the stability of DMRWDs. \label{fig:mrho}}
\end{figure}

An issue of concern is whether DMRWDs are stable against secular instability. When secular instability sets in, the star evolves until it reaches the point for dynamical instability, where it is unstable against gravitational collapse \citep{2003LRR.....6....3S}. As pointed out by \citet{1988ApJ...325..722F} and \citet{2013ApJ...762..117B}, a turning-point method could be used to identify the onset of secular instability:
\begin{equation}
    \left(\frac{\partial M}{\partial \rho_{c}}\right)_{J} = 0.
\end{equation}
Here, the NM central density is $\rho_{c}$. It has been shown that rotating WDs in the Newtonian regime are stable against secular instability \citep{1968ApJ...151.1089O, 2013ApJ...762..117B}. We show the total mass against $\rho_{c}$ for DMRWDs for different $\epsilon$ in Figure \ref{fig:mrho}, together with constant $J$ contour lines. We do not observe any local extremum for constant $J$ lines, and thus we can conclude that DMRWDs are also stable against secular instability in the Newtonian framework. We also arrive at the same conclusion for differentially rotating models. We note that general-relativistic effects would destabilize static and rotating WDs \citep{2013ApJ...762..117B}. Therefore, changes to this set of results are expected in the general-relativistic framework. In particular, it would be interesting to investigate whether admixing DM would postpone or advance the critical point of instability or even create more than one instability point.


\section{Generalizing the Tidal Equation to the Two-Fluid Case} \label{sec:2feqs}
Radau's equation which gives the tidal love number can be obtained in an alternative way by considering the Poisson equation \citep{2008ApJ...677.1216H}: 
\begin{equation}
\label{eqn:newtonian}
H''(r) + \frac{2}{r}H'(r) - \left[\frac{6}{r^{2}} - 4\pi G\rho(r)\frac{d\rho(r)}{dP(r)}\right]H(r) = 0.
\end{equation}
Here, $H(r)$ represents the Eulerian change of the Newtonian gravitational potential \citep{2017MNRAS.472.4965Y}. A change of the variable $Y(r) = r\frac{H'(r)}{H(r)}$ transforms the second-order equation to a first-order equation \citep{2017MNRAS.472.4965Y}:
\begin{equation}
Y'(r) + \frac{Y(r)}{r} + \frac{Y(r)^{2}}{r} = \frac{6}{r} + \frac{4\pi r^{3}}{m(r)}\frac{d\rho(r)}{dr},
\end{equation}
where we have made use of the chain rule $\frac{d\rho(r)}{dP(r)} = \frac{d\rho(r)}{dr}\frac{dr}{dP(r)}$ and the hydrostatic equation $\frac{dp(r)}{dr} = -\frac{Gm(r)\rho(r)}{r^{2}}$. We can define $y(r) = Y(r) - \frac{4\pi r^{3}}{m(r)}$ to obtain \citep{2017MNRAS.472.4965Y}:
\begin{equation}
y'(r) + \left[\frac{1}{r} + \frac{8\pi r^{2}\rho(r)}{m(r)}\right]y(r) + \frac{y(r)^{2}}{r} = \frac{6}{r} - \frac{16\pi r^{2}\rho(r)}{m(r)}.
\end{equation}
Rearranging terms, and denoting $D(r) = \frac{4\pi r^{3}\rho(r)}{3m(r)}$, we have:
\begin{equation}
ry'(r) + y(r)^{2} + y + 6D(r)(y(r) + 2) - 6 = 0,
\end{equation}
This is to be solved with initial condition $y(0) = -1 $, and the tidal deformability is given as $\tilde{k} = \frac{2 - y(R)}{2[3 + y(R)]}$, where $R$ is the stellar radius. By inspection, we make a substitution $y = \eta - 1$ to transform the equation as:
\begin{equation}
r\eta(r)' + \eta(r)(\eta(r) - 1) + 6D(r)(\eta(r) + 1) - 6 = 0,
\end{equation}
for which we recover Radau's equation with $\eta = \tilde{\eta}$. We notice that there are two-fluid generalizations to calculate the tidal love number of hybrid stars under the general relativistic framework \citep{PhysRevD.105.123010, zhang2020gw170817}. Therefore, the key to justifying our naive generalization of Equation \ref{eqn:randau} to the two-fluid situation is to take the Newtonian limit of the relativistic version of Equation \ref{eqn:newtonian} that has been established for the two-fluid system. The relativistic version of the equation is \citep{2008ApJ...677.1216H}:
\begin{equation}
\label{eqn:relativity}
\begin{aligned}
rY'(r) + Y(r)^{2} + Y(r)e^{\lambda(r)}[1 + 4\pi r^{2}(p(r) - \rho(r))] \\ + r^{2}Q(r) = 0, \\
Q(r) = 4\pi e^{\lambda(r)}\left[5\rho(r) + 9p(r) + \frac{\rho(r) + p(r)}{dp/d\rho}\right] - \\ 6\frac{e^{\lambda(r)}}{r^{2}} - (\nu'(r))^{2},
\end{aligned}
\end{equation}
where we have adopted geometric units $c = G = 1$. Here, $\lambda(r)$ and $\nu(r)$ are related to the metric elements. The term $\frac{\rho(r) + p(r)}{dp/d\rho}$ should be treated carefully in the two-fluid case. Fortunately, it can be decomposed into the contributions from NM and DM if they do not interact with each other \citep{PhysRevD.105.123010, zhang2020gw170817}:
\begin{equation}
\frac{\rho(r) + p(r)}{dp/d\rho} = \frac{\rho_{1}(r) + p_{1}(r)}{dp_{1}/d\rho_{1}} +\frac{\rho_{2}(r) + p_{2}(r)}{dp_{2}/d\rho_{2}}.
\end{equation}
In the Newtonian limit $e^{\lambda(r)} \approx 1$, $\nu'(r) \approx 0$ and $p \lll \rho$, we have:
\begin{equation}
Q(r) \approx 4\pi \left[\rho_{1}(r)\frac{d\rho_{1}(r)}{dp_{1}} + \rho_{2}(r)\frac{d\rho_{2}(r)}{dp_{2}}\right] - \frac{6}{r^{2}}.
\end{equation}
We make use of the hydrostatic equation $\frac{dp_{i}(r)}{dr} = -\frac{Gm(r)\rho_{i}(r)}{r^{2}}$ to obtain:
\begin{equation}
Q(r) \approx -4\pi \frac{r^{2}}{m(r)}\left[\frac{d\rho_{1}(r)}{dr} + \frac{d\rho_{2}(r)}{dr}\right] - \frac{6}{r^{2}}.
\end{equation}
Since the densities of DM and NM can be added as scalars, we substitute this expression into Equation \ref{eqn:relativity} and use $4\pi r^{2}(p(r) - \rho(r)) \approx 0$ to get:
\begin{equation}
	rY'(r) + Y(r)^{2} + Y(r) - \left(4\pi \frac{r^{4}}{m(r)}\frac{d\rho(r)}{dr} + 6\right) = 0,
\end{equation}
which is just Equation \ref{eqn:newtonian} with $\rho = \rho_{1} + \rho_{2}$ and $m(r)$ the total enclosed mass.

\section{The Formation of DMRWD} \label{sec:dmawdcreate}
We consider a scenario similar to that presented in \citet{2013PhRvD..87l3506L}, where the star is born with an inherent admixture of DM, contributing an extra gravitational force to the zero-age main-sequence star. We assume a spherically symmetric cloud of NM and DM having constant densities $\rho_{1}$ and $\rho_{2}$, respectively. Their individual radii could be computed by $R = (3M/4\pi \rho)^{3}$. In particular, we consider the situation with the DM radius $R_{1}$ being larger than that of the NM, $R_{2}$. The total energy $E$ (gravitational + kinetic) is:
\begin{equation}
\begin{aligned}
	E = -\left(\frac{3}{5}\frac{GM_{1}^{2}}{R_{1}} + \frac{3}{5}\frac{GM_{2}^{2}}{R_{2}} + \frac{3}{2}\frac{GM_{1}M_{2}}{R_{1}} - \frac{3}{10}\frac{GM_{1}^{2}R_{1}^{2}}{R_{1}^{3}}\right) \\
	+ \frac{3}{2}NkT + \frac{1}{2}M_{1}v_{1}^{2}.
\end{aligned}
\end{equation}
Here, $v_{1}$ is the DM thermal velocity, $N = M_{2}/m_{\text{H}}$ is the total number of NM nuclei, and $m_{H}$ is the molecular mass of hydrogen. Furthermore, since we widely analyzed the model with $\epsilon = 0.2$ in the text, we assume $M_{1} \sim 1.4\times0.2 \sim 0.3$ $M_{\odot}$, and $M_{2} \sim 10.0$ $M_{\odot}$. For a typical collapsing molecular cloud, we have $T \sim 150$ K and $\rho_{2} \sim 10^{8}m_{\text{H}}$ cm$^{-3}$, and hence $R_{2} = 3.05\times 10^{16}$ cm is smaller than the Jeans radius. The maximum velocity of DM $v_{1\text{max}}$ for it to be bounded by the combined gravitational force is obtained by solving $E(R_{2})$ = 0. We find $v_{1\text{max}} \sim 7.81\times 10^{5}$ cm s$^{-1}$.  For a given $v_{1} < v_{1\text{max}}$, we would fix $R_{2}$ and vary $R_{1}$ to look for solution where $E < 0$. However, the most probable DM speed (assuming a Maxwell distribution) is $v_{\text{p}1} \sim 10^{7}$ cm s$^{-1}$. To take this into account, the bounded DM fraction is given by $f$:
\begin{align}
	f = \frac{\int_{0}^{u_{1}}u^{2}\text{exp}(-u^{2})du}{\int_{0}^{\infty}u^{2}\text{exp}(-u^{2})du}.
\end{align}
Here, $u = v/v_{\text{p}1}$, and $u_{1} = v_{1}/v_{\text{p}1}$. We take a particular $v_{1} = 7.10\times 10^{5}$ cm s$^{-1}$, and give two sets of solutions in terms of $(R_{1}, \rho_{1})$ to show that the requirement of $E < 0$ could be satisfied: ($4.38\times10^{18}$ cm, 3539 GeV/cm$^{3}$) and ($3.05\times10^{16}$ cm, $1.05\times10^{10}$ GeV/cm$^{3}$). The required DM density in the first set of solutions is based on the state-of-the-art simulations, which showed that the DM density at the galactic bulge could be $\sim 3600$ GeV cm$^{-3}$ \citep{2014MNRAS.445.3133P}. The required DM density in the other set of solutions is much larger. However, such a value is possible near the galactic center, and values with a similar order of magnitude have been adopted in studying the effect of DM annihilation on main-sequence stars \citep{2006astro.ph..8535M, 2008ApJ...677L...1I}. In conclusion, our estimations considering the DM velocity dispersions show that it is possible to trap a DM of $0.3$ $M_{\odot}$ during the star-forming phase, provided that the molecular cloud is in the vicinity of the galactic center. Note that the DM and NM have different ambient densities, implying that they have different free-fall times, and in principle, the DM would not follow the trajectory of the NM.

\bibliography{sample631}{}
\bibliographystyle{aasjournal}

\end{document}